\documentclass[10pt,twocolumn,letterpaper]{article}

\usepackage{iccv}
\usepackage{times}
\usepackage{epsfig}
\usepackage{graphicx}
\usepackage{amsmath}
\usepackage{amssymb}

\usepackage{url}
\usepackage{bbding}
\usepackage{pifont}
\usepackage{enumitem}
\usepackage{multirow}
\usepackage{verbatim}
\usepackage{wrapfig}
\usepackage{subfigure}
\usepackage{setspace}
\usepackage{amsfonts}
\usepackage{algorithmic}
\usepackage{multicol}
\usepackage{xcolor,colortbl} 
\usepackage[ruled,vlined]{algorithm2e}

\definecolor{OliveGreen}{rgb}{0., 0., 1.0}
\usepackage[pagebackref=false,breaklinks=true,letterpaper=true,colorlinks,urlcolor = black,  citecolor = OliveGreen, bookmarks=false]{hyperref}

\usepackage[capitalize]{cleveref}

\iccvfinalcopy 

\def\iccvPaperID{2399} 

\ificcvfinal\fi

\begin{document}

\title{AutoMatch: A Large-scale Audio Beat Matching \\Benchmark for Boosting Deep Learning Assistant Video Editing}

\author{
Sen Pei$^{1,2}$, \;Jingya Yu$^1$, \;Qi Chen$^1$, \;Wozhou He$^1$\\
$^1$ByteDance Inc\\
$^2$NLPR, Institute of Automation, Chinese Academy of Sciences\\
{\tt\small{\{{peisen, yujingya.yjy, chenqi.02, hewozhou}\}@bytedance.com}}}

\maketitle
\ificcvfinal\thispagestyle{empty}\fi

\begin{abstract}
The explosion of short videos has dramatically reshaped the manners people socialize, yielding a new trend for daily sharing and access to the latest information. These rich video resources, on the one hand, benefited from the popularization of portable devices with cameras, but on the other, they can not be independent of the valuable editing work contributed by numerous video creators. In this paper, we investigate a novel and practical problem, namely audio beat matching (ABM), which aims to recommend the proper transition time stamps based on the background music. This technique helps to ease the labor-intensive work during video editing, saving energy for creators so that they can focus more on the creativity of video content. We formally define the ABM problem and its evaluation protocol. Meanwhile, a large-scale audio dataset, i.e., the AutoMatch with over 87k finely annotated background music, is presented to facilitate this newly opened research direction. To further lay solid foundations for the following study, we also propose a novel model termed BeatX to tackle this challenging task. Alongside, we creatively present the concept of \textbf{label scope}, which eliminates the data imbalance issues and assigns adaptive weights for the ground truth during the training procedure in one stop. Though plentiful short video platforms have flourished for a long time, the relevant research concerning this scenario is not sufficient, and to the best of our knowledge, AutoMatch is the first large-scale dataset to tackle the audio beat matching problem. We hope the released dataset and our competitive baseline can encourage more attention to this line of research. The dataset and codes will be made publicly available.
\end{abstract}

\section{Introduction}

\begin{figure}[t]
\begin{center}
\includegraphics[width=0.97\columnwidth]{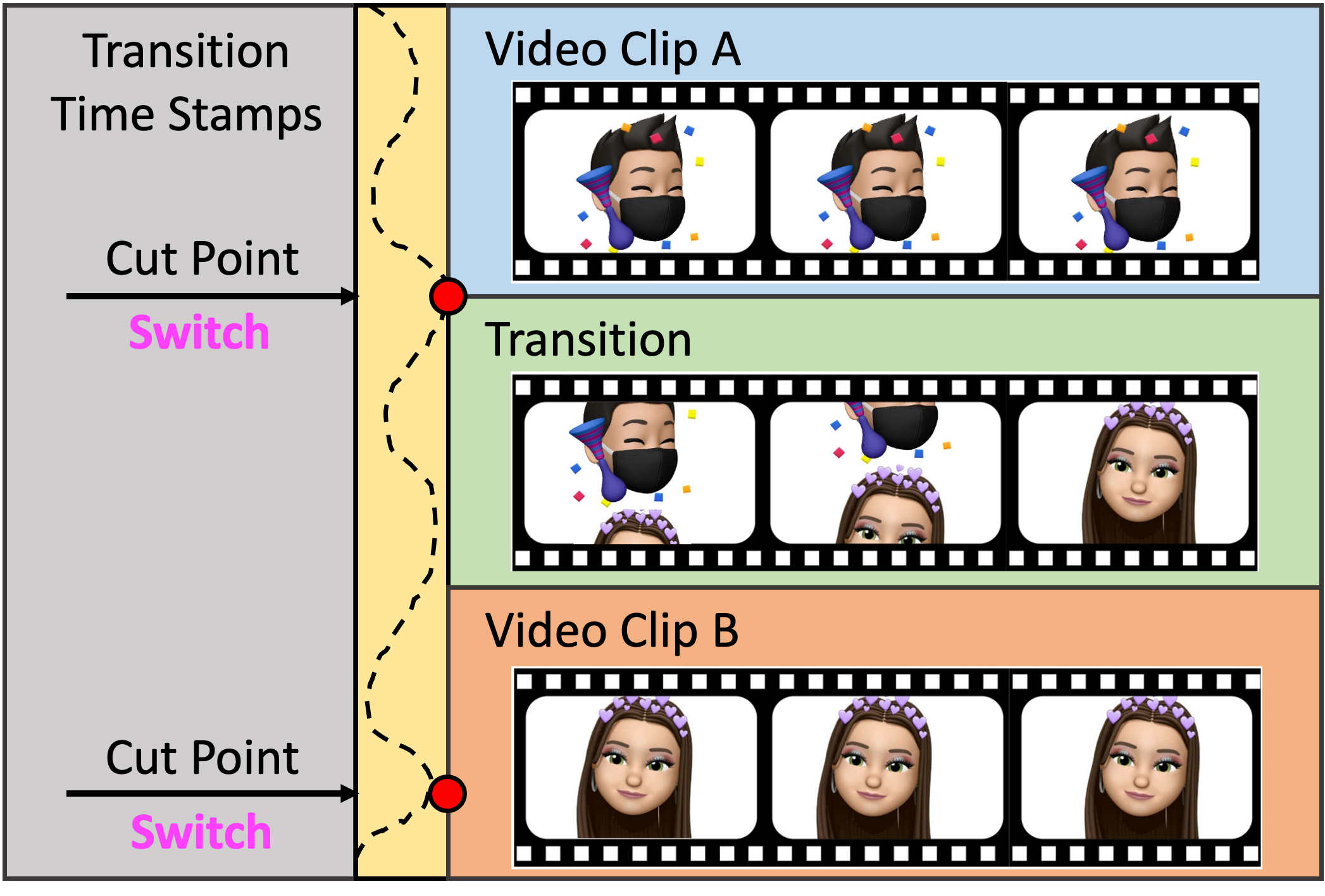}
\caption{An example of manual video editing. Each group of images above demonstrates a scene appearing in the video. The wave on the left is the corresponding audio information. The video creator assigns transition effects among the different video contexts. The transitions are placed at the proper location to match the beat of the audio. Without resorting to the pre-defined templates, the user has to locate the cut points of audio laboriously, which contributes less to the creativity of videos.}
\label{fig:bytefood}
\end{center}
\vskip -0.35in
\end{figure}

The research in deep learning has brought practical convenience to peoples' lives, for example, the bio-information authentication \cite{quantum_dot, ai_interaction, video_authenticity, deepface, maskface, adaface, synface}, the recommendation system \cite{pairwise, listwise, deepfm, deepctrTool, PLE, DIFM, deepLTR, EDCN, enhancing}, and the assistant robots \cite{constrained_control, home_assistant, health_care, riserobots, reachbot, bite_transfer, navigation}, to name a few. No matter how macro or specific the task is, the solution of a problem often leads to the explosive growth of the related applications, saving people from the heavy repetitive and non-technical operations. Inspired by many research relevant to big data and AI such as \cite{ai_good, ai_social, solar_energy, BankNote, bigdata}, we argue that during the era of big data, properly defining/modeling a practical task should be endowed with equal importance of solving the problem to some extent, since it is an effective way of leading AI for good. This principle has promoted numerous AI researchers to investigate and step close to the areas which are not commonly concerned, spreading the bonus of AI research broadly. 

In this paper, we maintain the original intention that technology should facilitate our life, and therefore, we focus on an area of great significance but lack of relevant research, \emph{i.e.}, detecting the potential cut points within the audio. Formally, the cut point indicates a location within the video where the content is suggested to change, and usually, a transition is inserted to connect the different contexts. We term this task as audio beat matching (ABM), which plays a vital role in video creation since the mismatch between the switching of video content and the rhythm of the background music will introduce an uncomfortable sense to the viewers. Based on the present, the editors finish most of the aforementioned ABM tasks manually during the video editing procedure, taking them a lot of time and energy. To empower video creators and promote automatic video editing, we notice and further investigate this unpopular direction, \emph{i.e.}, ABM, aiming to enable deep neural networks with human-like editing abilities.

To the best of our knowledge, existing work has developed the ability to detect the changing boundaries of audio using methods such as zero-cross rate \cite{zero_cross} and classic Fourier transform. Still, these capabilities can only provide limited benefits for video creators. For example, in a great many videos, we can effortlessly notice that the changing boundaries of music are not necessarily leading to transitions. This evidenced that the time stamps of the aforementioned cut points are highly relevant but do not entirely depend on the local changes of music, revealing that the video editors have to consider more than the local music boundaries, for example, the fashion of the entire music.

To help the deep models develop such human-like editing ability, we collect the first large-scale \textbf{audio to transitions} dataset, namely the \emph{AutoMatch}, which contains over 87k background music and their corresponding cut point time stamps. With these foundations, we further develop an algorithm termed \emph{BeatX} to automatically analyze the raw audio file and report the recommended transition locations using local-to-global features. This assistant technique will undertake some preliminary work during the video editing, enriching the styles of video creation. The main contributions of this paper are summarized in three aspects:

\begin{itemize}
\vspace{-0.5em}
\item We formally define the task of audio beat matching (ABM), including the evaluation protocol, which aims to help the video creators locate the proper transition time stamps based on the audio information. This technique will provide recommended cut points for numerous video editors, partly freeing them from the tedious pre-processing work. 

\vspace{-0.5em}
\item We collect the first large-scale dataset named \emph{AutoMatch} to facilitate the research of ABM. The \emph{AutoMatch} consists of over 87k finely annotated background music and the corresponding transition time stamps. This dataset helps the ABM to enjoy the benefits of deep learning schemes.

\vspace{-0.5em}
\item We propose \emph{BeatX}, which takes advantage of classic and deep learning techniques, achieving competitive detection performance. Besides, we present the label scope, eliminating the data imbalance issues and assigning adaptive weights for the loss. The extensive experiments and detailed ablations demonstrate the insights and effectiveness of our proposed \emph{BeatX}. We hope our work can inspire more researchers to focus more on unpopular but practical directions, promoting the convenience of life with the latest technology.  
\end{itemize}

\section{Related Work}
\label{related_work}
In this section, we give a brief overview of the most related research, including the audio representation learning, the automatic video editing, and the generic event boundary detection (GEBD). 

\noindent
\textbf{Audio Representation Learning.} Both in the age of handcrafted descriptors and the deep features, the design of representations determines the model's performance. In \cite{nature_audio}, the work considers the ensemble of several classic audio features, such as the Mel cepstrum \cite{mfc}, the pitch class profile \cite{pitch_class}, the cross-zero rate \cite{zcr1, zcr2}, and the short-term energy \cite{nature_audio}. All these features are designed manually and elaborately with considerable generalization ability. \cite{audio_survey} further points out the four stages of audio feature evolution, which are the time domain, the frequency domain, the ensemble of time-frequency domains, and the latest deep learning based features. In this paper, we adopt both the classic manually designed features and the learned deep representations and report their corresponding effects. 

\noindent
\textbf{Automatic Video Editing.} This technique completes a lot of pre-processing work for video editors, which is one of the representatives of deep learning based assistant tools. AutoTransition proposed in \cite{auto_transition} aims to recommend the transitions between two subsequent video clips given their visual and audio information. This work also formally defines the video transition recommendation (VTR) task and its corresponding evaluation protocol, reducing the difficulty of transition selection for non-professionals. Non-linear editing transfer \cite{editing_transfer} investigates how to apply the editing manner of the source videos to the target. Text2live \cite{text2live} develops the ability to edit images or videos using the input texts, lowering the threshold of video editing. MovieCuts \cite{moviecuts} models the multi-modal information to recognize the cut type, which unleashes new experiences in the video editing industry. CMVE \cite{human_editing} leverages both the visual and textual metadata and can highlight the key objects relevant to the main storyline appearing in the raw videos, helping the editors to locate their concerned clips quickly. Talk-head video editing \cite{text_talk_editing} devotes to helping the editors change the speech content or remove filler words within the interview videos. Above, we just list several representative examples of deep learning assistant tools in video editing. All these mentioned methods either have tackled some practical issues which ease the video creators' working load or open a new research direction that has not been sufficiently investigated. Keeping up with these inspirations, in this work, we notice an unpopular but practical direction, the audio beat matching (ABM), which aims to find the potential transition locations based on the background music, easing a part of the work the editors have had to do previously.

\noindent
\textbf{Generic Event Boundary Detection.} GEBD has been long researched to locate the taxonomy-free event boundaries in a raw video, breaking it into several subsequent temporal clips. Techniques in \cite{gebd_report, gebd_winning, gebd_discerning, gebd_benchmark} build the contexts as a group of frames in the fixed length, distinguishing whether each frame belongs to boundaries or not separately. Unlike these existing arts, SC-Transformer proposed in \cite{sc-transformer, sc-transformer++} presents SPoS module to efficiently generate the contexts for each frame and applies temporal fusion resorting to the transformer encoder, empowering feature interaction along the temporal axis. Concerning the audio beat matching (ABM) task, the location of transitions also can be regarded as the boundaries of background music. However, unlike the GEBD, which can be identified using local contexts, the boundaries appearing in ABM are not solely determined by the local rhythm (\emph{i.e.}, contexts) which means the global information of background music can dramatically affect the recommendation performance.

\section{Preliminaries}
In this section, we detail the background of the proposed audio beat matching (ABM) task, including the problem definition and the evaluation protocol.

\subsection{Problem Definition}
\label{definition}
Given a whole audio input depicted as $x$, we split it into several subsequent clips with the basic time unit, \emph{i.e.}, $x=\{x_1, x_2,...,x_N\}$. The time unit indicates the duration of each audio clip $x_k$. Concerning that the original duration of audio $x$ may not be divisible by the time unit, we pad at the end of $x$ using zeros, which means the whole duration of all audio clips is not less than the duration of the raw input $x$. Correspondingly, the target label of $x$ is demonstrated as $y$, which is a binary vector with $N$ values. For example, $y_i=1$ reveals it is suggested to change the video content and assign a proper transition after the \emph{i-th} time unit. The task of ABM aims to build a predictor $f(\cdot)$, which takes the audio $x$ as the input and gives out a predicted vector $\hat{y}$ in $N$ dimension. The prediction $\hat{y}$ is expected to be as similar as possible to the ground truth $y$.

\subsection{Evaluation Protocol}
\label{protocol}
We use recall (\emph{R}) and precision (\emph{P}) to evaluate the performance of the predictor in ABM. Beforehand, we first give the definition of \emph{hit@k}. We call the prediction $\hat{y}_i$ \textbf{hits} the targets $y$ if both $\hat{y}_i$ and $y_i$ equal to 1, which means a transition should be inserted after the \emph{i-th} time unit. Furthermore, we relax the evaluation criteria. If the prediction $\hat{y}_i$ and one of the targets between $\{y_{i-k}, y_{i-k+1},...,y_i,...,y_{i+k-1},y_{i+k}\}$ equal to 1, we call this situation as \emph{hit@k}. Apparently, the conventional prediction performance is calculated under the setting of \emph{hit@0}. With these notations and definitions, we build the \emph{P@k} (precision) and \emph{R@k} (recall) as follows:
\begin{equation}
\centering
P\emph{@}k=\frac{hit\emph{@}k(y, \hat{y_i})}{\sum_{i=1}^{N}\hat{y}_i}
\;\;\;\;
R\emph{@}k=\frac{hit\emph{@}k(y, \hat{y_i})}{\sum_{i=1}^{N}y_i}
\label{eq:p_r}
\end{equation}
where \emph{hit@k($y,\hat{y}$)} indicates the number of overlapped values between $y$ and $\hat{y}$ under the setting of \emph{hit@k}. Based on the above protocol in Eq.\ref{eq:p_r}, one can effortlessly calculate other metrics such as F1-Score, AUPR, and AUROC.

\section{The \emph{AutoMatch} Dataset}
In this section, we detail the process of data collection, quality control, and data safety. Following that, we demonstrate the statistical information of \emph{AutoMatch}. Note that the audio is extracted from the collected online videos.

\begin{figure*}[t]
\centering
\subfigure[]{
\begin{minipage}[t]{0.331\textwidth}
\centering
\includegraphics[width=2.14in]{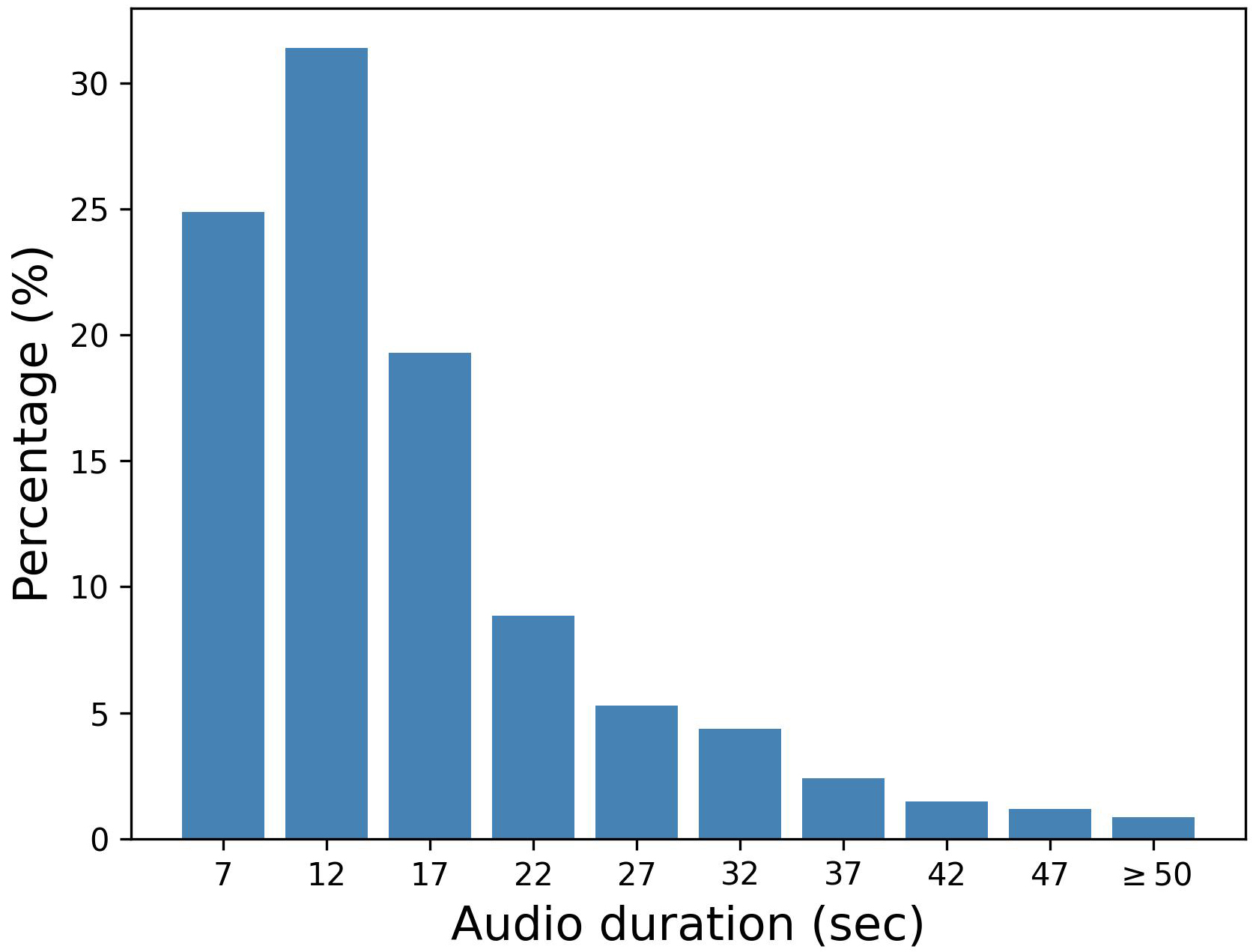}
\end{minipage}%
}%
\subfigure[]{
\begin{minipage}[t]{0.331\textwidth}
\centering
\includegraphics[width=2.14in]{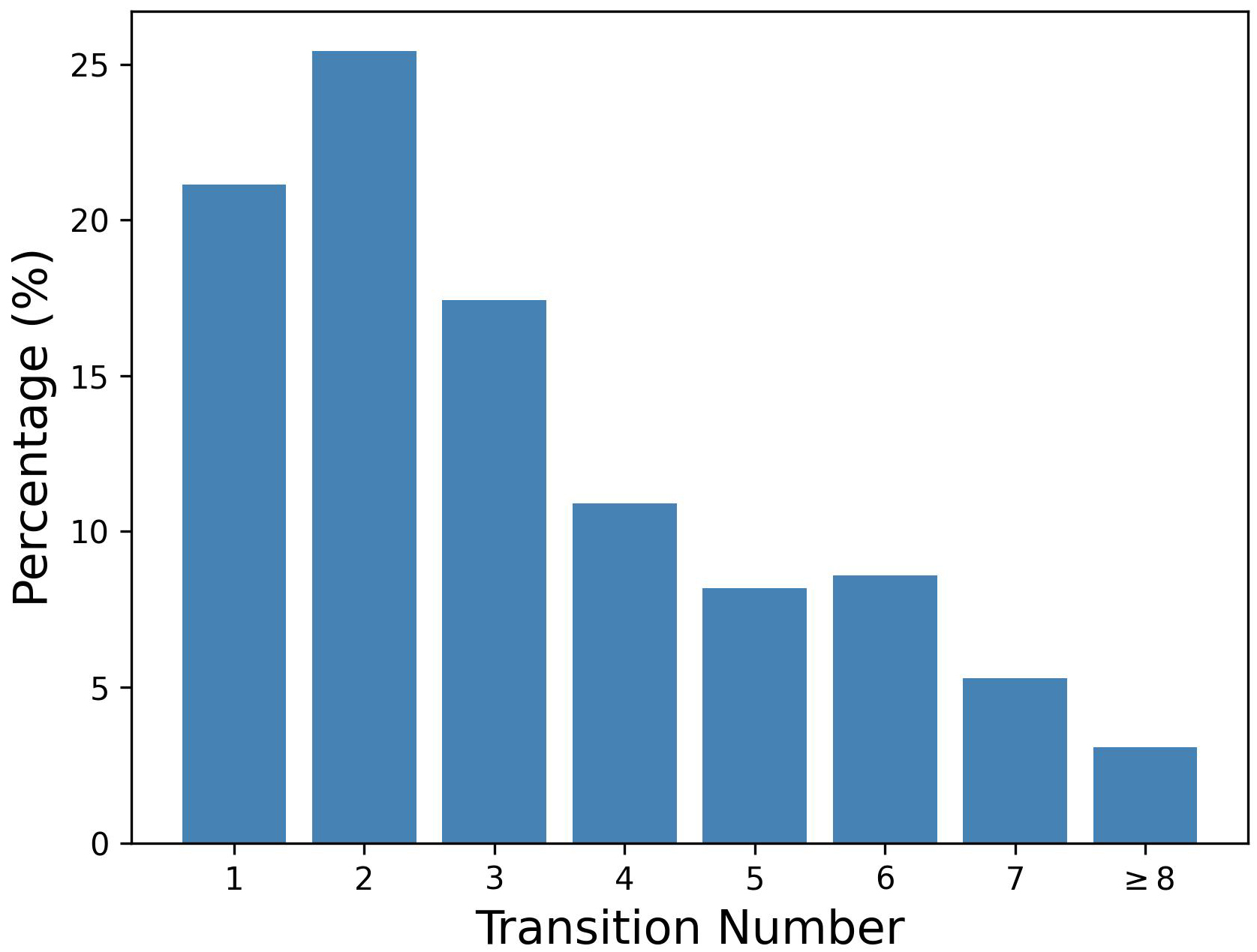}
\end{minipage}%
}%
\subfigure[]{
\begin{minipage}[t]{0.331\textwidth}
\centering
\includegraphics[width=2.14in]{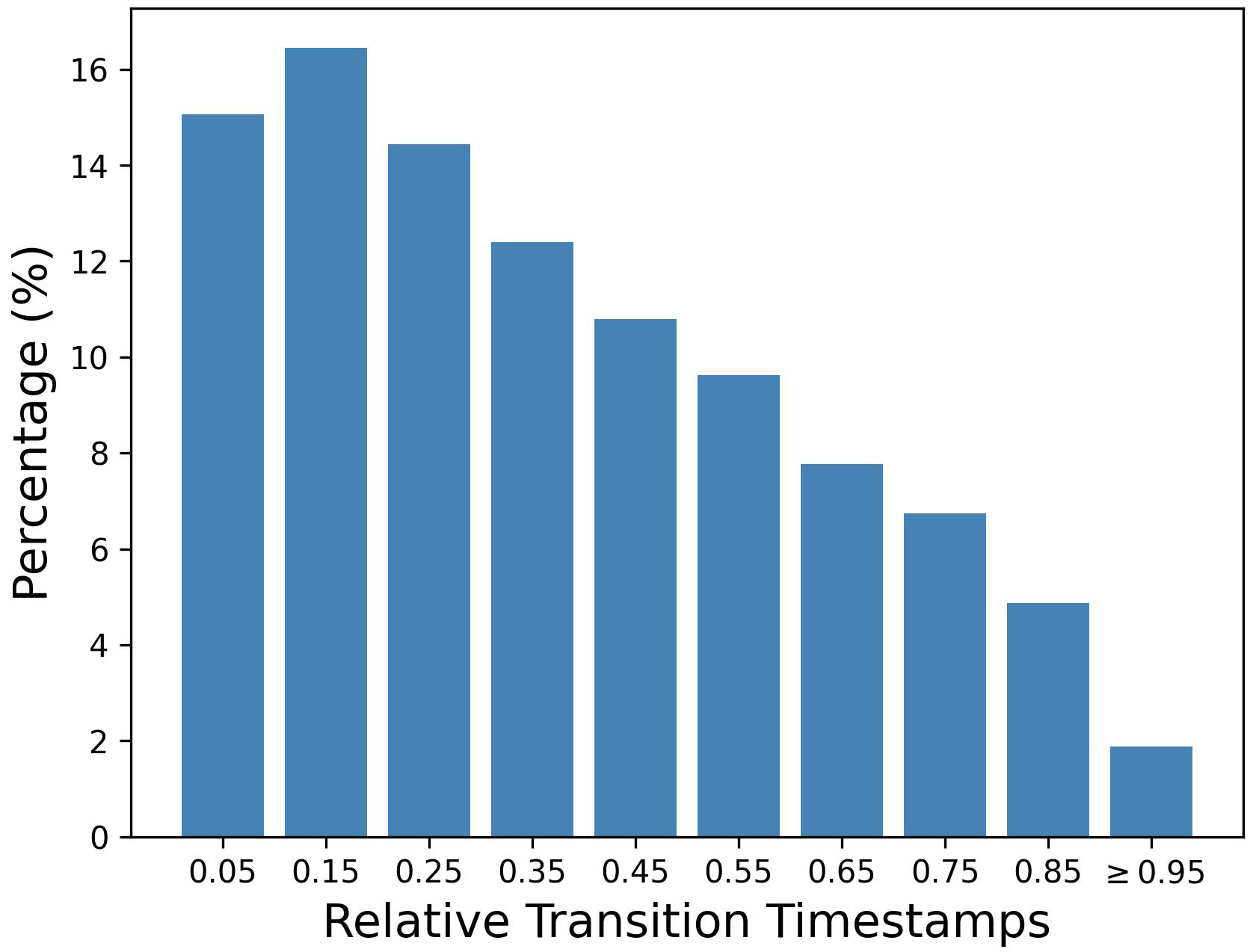}
\end{minipage}%
}%
\centering
\caption{Statistical results of the proposed \emph{AutoMatch}. (a) shows the distribution of audio duration. (b) illustrates the number of transitions (\emph{i.e.}, cut points) within each background music. We can see that most background music is assigned with less than 8 transitions. (c) shows the relative time stamp of each transition appearing in the corresponding videos. It is obvious that the distribution of the relative transition time stamps is evenly distributed instead of gathering at some specific regions, which means the annotations of \emph{AutoMatch} are rich in diversity and thus challenging to learn. These complex ground truths are helpful to verify whether the model has captured effective representations or just remembered the pattern of this specific dataset.}
\label{fig:statistics}
\vskip -0.1in
\end{figure*}

\subsection{Data Collection}
\label{data collection}
Collecting such a large dataset is undoubtedly a challenging task. To increase efficiency and avoid complex procedures in authority, we constrain that only the videos satisfy the following rules can be treated as the candidates of \emph{AutoMatch}: 1). The videos are open-sourced and publicly available by all users on the platforms, and the video creators allow the users from the Internet to perform editing or re-creation based on their original copy. This will help us to get fully authorized by the video uploaders/owners conveniently; 2). The contents, including but not limited to the images, the topics, and the audio, \emph{etc}, used in the videos are healthy and positive, \emph{i.e.}, no x-rated frames, no uncomfortable music, and no personal information, to name a few. This constraint plays a vital role in making the \emph{AutoMatch} free of ethical concerns or issues; 3). No copyright issues. The background music used in the videos does not infringe on others' copyright. This is relatively easy to guarantee since the auditing mechanism of short video-sharing platforms is strict. The background music of the video which meets the above criteria is picked as one of the candidates, and we further check whether its transitions are highly related to the music using the tags annotated by the video creators in case to reserve the rhythmic videos while ruling out the noisy data, \emph{i.e.}, the transitions show rare relevance with the audio. 

\subsection{Data Statistics}
We analyze the distribution of audio duration, the number of transitions within each audio, and the relevant time stamps of each transition. The statistical results are depicted in Figure \ref{fig:statistics}. From Figure \ref{fig:statistics} (a), we can notice that extremely long audio with over one minute is ruled out if it has less than ten corresponding cut points. These audios usually have a very soothing melody, and the corresponding cut points distribute temporal evenly, contributing less to the learning for discriminative features. Figure \ref{fig:statistics} (b) demonstrates the distribution of cut points quantity within each audio. Since we rule out the extremely long and smoothing music in \emph{AutoMatch}, most of the audios have less than eight transitions/cut points. Figure \ref{fig:statistics} (c) tells that the cut points distribute evenly within the audios, eliminating the shortcut learning \cite{shortcuts, overcome_shortcut} that deep neural networks directly capture the specific pattern of this dataset.

\subsection{Data Safety}
We promise the safety of data in a two-step manner. First, we have introduced the checking mechanism during the data collection procedure (cf. Section \ref{data collection}). Besides, we also cooperate with the in-house legal/privacy staff to ensure that our operations during the research meet the requirements of the related law, including data collection, data processing, and data usage, \emph{etc}. We hire no workers or human participants during our research. Thus we do not have to prepare the Informed Consent Form (ICF) in advance. All data used in \emph{AutoMatch} has been examined by the in-house legal department, freeing it from potential legal/ethical risks and copyright issues.

\section{Methods}

In this section, we detail the motivations beyond our model architecture design, the construction of loss function, and the reproduction procedure.

\subsection{Architecture Design}
We focus on three fundamental problems in designing the model: 1). Use manually designed features or adaptively learned representations. As introduced in Section \ref{related_work}, the deep learning models can be empowered with the leverage of elaborately designed descriptors such as Mel spectrum \cite{mfc} and short-term energy \cite{nature_audio}, we accept this suggestion in our designing, performing fusion among the traditional and deep features; 2). How to capture the relations among audio neighbors, \emph{i.e.}, the adjacent audio clips. The main challenges of this topic are in two aspects, selecting a proper window size to perform local attention and generating the structural context \cite{sc-transformer} for each clip efficiently during the training; 3). How to perform global attention in a linear complexity with respect to the audio duration. As observed in many research \cite{lowrank, glu, efficient_att, linformer, roformer, reformer, trans_rnn}, though the transformer has yielded impressive performance on many complicated tasks, training or deploying it into practical scenarios is prohibitively expensive since it incurs a complexity of $O(N^2)$ with respect to the sequence length $N$. Since the ABM serves practical and real-time scenarios, the inference cost appears particularly critical.

\begin{figure*}[t]
\begin{center}
\includegraphics[width=\textwidth]{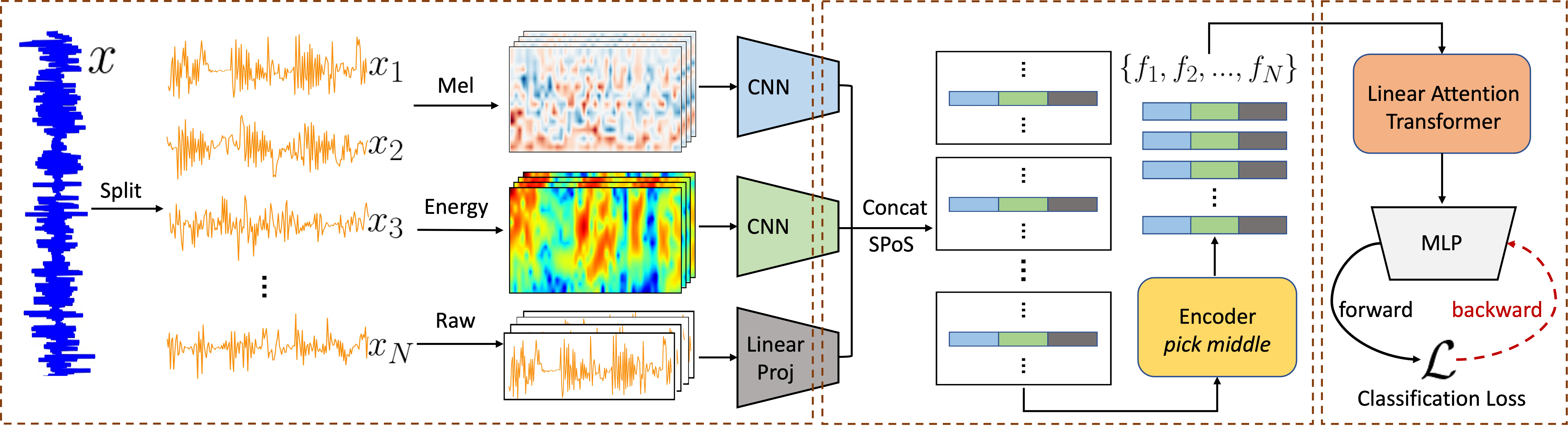}
\caption{The proposed \emph{BeatX} framework. In the above figure, we highlight the aforementioned three components of \emph{BeatX} using the dashed box. From left to right, they are Time-wise Feature Extraction (TFE), Local Contexts Generation (LCG), and Linear Global Fusion (LGF). The input signal is split into subsequent audio clips whose duration equals the pre-defined time unit. The Mel cepstrum and short-term energy are adopted as the manual descriptors, alongside, the raw audio signal will be processed with a linear projection layer. All these features are concatenated as the intermediate representations. We resort to SPoS \cite{sc-transformer} for generating structural contexts of each audio unit $x_k$. Following that, a conventional encoder and a linear attention transformer (LinFormer) \cite{linformer} are responsible for completing the local and global feature fusion. The overall pipeline is built as a classification problem and optimized using the back-propagation technique.}
\label{fig:framework}
\end{center}
\vskip -0.25in
\end{figure*}

To summarize, we argue that a suitable model for the ABM task should employ generalizable audio representations, obtain structural contexts efficiently, and perform the global attention across the whole audio in nearly linear complexity with respect to the audio duration. To deal with the above obstacles, we design \emph{BeatX}, which takes both the raw audio signal and manually designed descriptors as input, resorts to SPoS \cite{sc-transformer} for generating structural contexts quickly, and fuses local features using both training and deploying economical LinFormer \cite{linformer}. The overall framework of our proposed \emph{BeatX} is depicted in Figure \ref{fig:framework} at length. To facilitate the description, we split the \emph{BeatX} into three components as discussed above, which are Time-wise Feature Extraction (TFE), Local Contexts Generation (LCG), and Linear Global Fusion (LGF). Following the pipeline shown in Figure \ref{fig:framework}, we formalize each module and define the notations for building loss functions.

\noindent
\textbf{Time-wise Feature Extraction (TFE).} As depicted in Section \ref{definition}, given the raw audio input $x$, we first split it into $N$ subsequent audio clips namely $\{x_1,x_2,x_3,...,x_N\}$. The duration of each audio equals the pre-defined time unit. No overlap is introduced among these sub-clips. To leverage the semantic information of manually designed descriptors, we input both the Mel cepstrum and short-term energy along with the raw signal. The CNN depicted in the above figure is the first two blocks of ResNet-18 \cite{resnet}. The raw signal is projected with a linear layer activated by ReLU. The output of the TFE module is the concatenation of the above feature, \emph{i.e.}, $g_i=M(x_i)\circ E(x_i)\circ L(x_i)$, where $M(\cdot)$, $E(\cdot)$, and $L(\cdot)$ indicate the Mel cepstrum, short-term energy, and linear projection, respectively. We use $\circ$ to represent the concatenation operation. No temporal interactions are attended among the local features in TFE.

\noindent
\textbf{Local Contexts Generation (LCG).} This module aims to efficiently generate structural contexts for each time unit (\emph{i.e.}, $x_i$). Given input $x=\{x_1,x_2,x_3,...,x_N\}$ and local window size $k$, LCG yields contexts of $x_i$ as $\{x_{i-k},x_{i-k+1},...,x_i,...,x_{i+k-1},x_{i+k}\}$ with length $2k+1$. Contexts with insufficient duration, for example, the top and last $k$ clips, are padded with zero vectors. Supposing the split audio is demonstrated as $\{x_1,x_2,x_3,...,x_N\}\in \mathbb{R}^{N\times C}$ where $C$ is the channel dimension, then the structural output of LCG is $\{c_1,c_2,c_3,...,c_N\} \in \mathbb{R}^{N\times k\times C}$, where $c_i\in \mathbb{R}^{k\times C}$ indicates the local context of audio $x_i$. SPoS \cite{sc-transformer} achieves the requirements mentioned above using well-designed array arrangements. Since this is more like a kind of engineering scheme, we do not detail much here. For jointly attending the information at different time stamps, each structured context $c_i$ is fused by the multi-head encoder. We only pick the middle token of each context $c_i$, and the output is notated as $f_i\in \mathbb{R}^C$. The group of structured features $\{f_1,f_2,f_3,...,f_N\}$ distribute in the space of $\mathbb{R}^{N\times C}$, which is identical as $\{x_1,x_2,x_3,...,x_N\}$.

\noindent
\textbf{Linear Global Fusion (LGF).} Generally, the main efficiency bottleneck of transformer architecture is incurred by the self-attention mechanism, since the representation of each token has to measure the correlation with all other tokens from the previous layer. Briefly, given $Q$, $K$, and $V$ ($\mathbb{R}^{N\times d}$) as the inputs, self-attention aims to learn a group of weights $W^Q$, $W^K$, and $W^V$ ($\mathbb{R}^{d\times d}$) to retain the long-term information across the sequence, and the output is built as:
\begin{equation}
h=\texttt{softmax}[\frac{QW^Q(KW^K)^T}{\sqrt{d}}]VW^V=PVW^V
\end{equation}

\noindent
where $d$ is the hidden dimension of the projection space yielded by $W^K$ and $P$ is the equivalent context mapping matrix. We do not distinguish the dimension of features before and after 
 the mapping of $W^Q$, $W^K$, and $W^V$. Calculating $P$ requires $O(N^2)$ complexity in both time and space since the sequence length of $Q$, $K$, and $V$ are $N$. LinFormer \cite{linformer} argues and verifies the mapping matrix $P$ of self-attention is low-rank and thus can be decomposed into multiple smaller attentions through linear projections. For example, if we pick $E$ and $F$ ($\mathbb{R}^{p\times N}$, $p\ll N$) as the linear projections performed on $K$ and $V$, we can built the approximate $\hat{h}$ as follows:
\begin{equation}
h\approx \hat{h}=\texttt{softmax}[\frac{QW^Q(E\cdot KW^K)^T}{\sqrt{d}}]F\cdot VW^V
\end{equation}

\noindent
Note that $p$ is the intermediate dimension of the linear projection. In the equation above, multiplying $QW^Q$ and $(E\cdot KW^Q)^T$ only incurs a complexity of $O(N)$ since both $p$ and $d$ are constants. This operation significantly mitigates the complexity of self-attention from $O(N^2)$ to $O(N)$, such that $p\ll N$. We adopt this principle when designing our global fusion module. The introduced performance drop is evaluated in LinFormer \cite{linformer}.

\subsection{Loss Functions}
We tackle two critical concerns in this section: 1). how to balance the positive and negative data since only very few time stamps are annotated as the proper transition locations, which means the negative data are far more than the positive; 2). how to measure the significance of loss incurred by the model's prediction since a short time offset (\emph{e.g.}, only a time unit) of the prediction is acceptable to some extent. For simplicity, the label of input $x$ is depicted as $y=\{y_1,y_2,y_3,...,y_N\}$ since $x$ is split into $N$ subsequent clips/time units. $y_i=1$ indicates a transition should be inserted immediately after the \emph{i-th} clip. 

With these auxiliary notations, we solve the above issues using \textbf{label scope} $s_i(k)$, which measures the effect of the \textbf{positive} label $y_i$ yielding at the \emph{k-th} time unit. Intuitively, the influence of $y_i$ decreases with the distance. For example, the label of the first time unit incurs marginal influence on the prediction of the last time unit. Empirically, the label scope $s_i(k)$ is built as a Gaussian function:
\begin{equation}
s_i(k)=\exp[-\frac{(i-k)^2}{2\sigma ^2}]
\label{eq:gaussian}
\end{equation}

\noindent
where $\sigma$ is the radius of the Gaussian distribution. With Eq. \ref{eq:gaussian}, we have solved the second problem posed at the beginning of this section. Furthermore, we use the hyper-parameter $\sigma$ to control the balance between the positive and negative data. Obviously, greater $\sigma$ emphases more on the positive data since the Gaussian distribution is stretched to be wider along the temporal axis. Moreover, the $\sigma$ is dataset-dependent since it should be determined based on the positive/negative label distribution of the specific dataset. To acquire the proper value of $\sigma$, we resort to the ablations performed in Section \ref{ablation_scope}. Presently, we just pick $\hat{\sigma}$ as the most appropriate fit of $\sigma$, and the label scope is refined as $s_i(k, \hat{\sigma})$. With these notations, given the input $x$ with $N$ clips and the corresponding target $y$, we obtain the overall importance of the \emph{k-th} time unit as following:
\begin{equation}
S(k, \hat{\sigma})=\frac{1}{Z}\sum_{i=1}^{N}s_i(k, \hat{\sigma}) \cdot y_i
\label{eq:gaussian}
\end{equation}

\noindent
where $Z$ is a normalization factor, which guarantees the maximum of importance is not greater than 1. $k$ starts from 1 to $N$. $S(k, \hat{\sigma})$ is treated as the label scope mask, which balances the positive/negative data and measures the significance of each time unit. Supposing the prediction of \emph{BeatX} is depicted as $P=\{p_1,p_2,p_3,...,p_N\}$, we formalize the complete loss objective in the following manner:
\begin{equation}
\mathcal{L}=-\frac{1}{N}\sum_{i=1}^{N}S(i, \hat{\sigma})\cdot[y_i\log(p_i)+(1-y_i)\log(1-p_i)]
\label{eq:loss}
\end{equation}

Specifically, we omit the mask $S(k, \hat{\sigma})$ for those videos with no transitions, since it always equals zero.

\subsection{Reproduction Statement}
We sincerely admire the open source circumstance built in the computer vision community. The detailed codes and dataset used in this work will be made publicly available. Besides, we also provide the script to help the users render their videos based on the image candidates and the raw audio. We design the interface in a user-friendly fashion, which takes the mp3/mp4/mov file as the input and extracts the audio information automatically. With the recommended time stamps of transitions from \emph{BeatX}, the rendering script will generate a demo combining the audio, images, transitions, and video effects integrally. Our demos have been attached in the \textbf{Supplementary Materials}.

\section{Experiments}
This section is responsible for demonstrating the details of our experiments, reporting the comparison results of \emph{BeatX}, presenting the effects of each component within \emph{BeatX}, and performing ablations on the hyper-parameters.

\setlength{\tabcolsep}{5pt}
\begin{table*}[t]
\begin{spacing}{1.47}
\begin{center}
\begin{tabular}{l|c|>{\columncolor{gray!25}}c|cc>{\columncolor{gray!25}}c|cc>{\columncolor{gray!25}}c|cc>{\columncolor{gray!25}}c}
\hline
\multicolumn{1}{c|}{\emph{Methods}} & \emph{Params} & \multicolumn{1}{c|}{\emph{ACC}} & \multicolumn{3}{c|}{\emph{hit@0}} & \multicolumn{3}{c|}{\emph{hit@1}} & \multicolumn{3}{c}{\emph{hit@2}} \\ \cline{4-12} 
\multicolumn{1}{c|}{} & &\multicolumn{1}{c|}{} & \multicolumn{1}{c}{\emph{P}} & \emph{R} & \multicolumn{1}{c|}{\emph{F1}} & \multicolumn{1}{c}{\emph{P}} & \multicolumn{1}{c}{\emph{R}} & \multicolumn{1}{c|}{\emph{F1}} & \multicolumn{1}{c}{\emph{P}} & \multicolumn{1}{c}{\emph{R}} & \multicolumn{1}{c}{\emph{F1}}\\ \hline
\emph{Linear}  & 5.44M & 75.97 & 65.46 & 49.26  & 53.82 & 81.77 &  52.20 & 63.16 & 88.43 &  65.74 &  71.60 \\
\emph{CNN-1d}  & 5.54M & 75.86 & 67.41 &  46.38 & 53.50 & 83.88 & 48.38 & 60.13 & 87.04 & 64.52 & 70.93 \\
\emph{Encoder} & 5.64M & \underline{77.12} & 67.10 & 47.29 & \underline{54.20} & 85.91 & 64.35 & \underline{73.24} & 88.27 &  71.30 & \underline{79.82} \\ \hline
\emph{BeatX} & 5.10M & \textbf{79.95\textcolor{red}{$_{+2.83}$}} & 67.21 & 60.36 & \textbf{61.58\textcolor{red}{$_{+7.38}$}} & 86.40 & 77.66 & \textbf{79.11\textcolor{red}{$_{+5.87}$}} & 91.23 & 80.76 & \textbf{83.08\textcolor{red}{$_{+3.26}$}} \\ \hline
\end{tabular}
\end{center}
\end{spacing}
\vskip -0.09in
\caption{Comparable results with previous arts. The detailed definition of \emph{hit@k} can be found in Section \ref{protocol}. The depicted overall metrics are obtained by averaging the audio-wise prediction results. We keep the comparable methods to enjoy similar training complexity. Noting that the \emph{hit@k} can also be efficiently addressed using SPoS \cite{sc-transformer}, we detail this in the codes of our attached Supplementary Materials. We highlight the best performance using boldface, and the closely followed methods are marked using underline.}
\label{tab:main}
\vskip -0.1in
\end{table*}

\subsection{Experimental Setup}
\noindent
\textbf{Comparable methods}.Performing comparison with previous methods is non-trivial since no sufficient research is proposed in this direction. To evidence the effectiveness of our proposed \emph{BeatX}, we design several comparable baselines, which are the Linear classifier, the one-dimensional CNN, and the self-attention based encoder. We keep the principle that all comparable methods enjoy similar computation complexity, \emph{i.e.}, the trainable parameters. The fully connected classifier (Linear in Table \ref{tab:main}) is built to have four linear layers activated by ReLU. The self-attention based encoder has 8 heads per layer and 4 layers in total. The feed-forward (FFN) dimension of the encoder is set to 512. The convolutional model employs four stacked one-dimensional convolution layers, following that, the intermediate features are projected using two subsequent fully connected layers. In \emph{BeatX}, the feature dimension of the Mel cepstrum, the short-term energy, and the raw signal is 128. The encoder used in LCG has 2 attention layers, each has 4 heads. Only a single attention layer with 4 heads is used in the LGF.

\noindent
\textbf{Details of the training and inference procedure}. All comparable methods are trained in a total of 300 epochs, the time unit is set to 0.1 seconds by default. The learning rate starts from 2e-4 and halves every 20 epochs. The \texttt{AdamW} is adopted as the optimizer. No pre-trained weights are employed in our experiments, which means all schemes are trained from scratch. We split 7,000 of the dataset for validation and report the best metric on them. The overall performance is obtained by averaging the audio-wise result. The window size (\emph{i.e.}, the semi-duration $k$ of the local context in LCG) is set to 5. All experiments are performed with 8 NVIDIA V100 \footnote{https://www.nvidia.com/en-sg/data-center/v100/} GPUs.


\subsection{Main Evaluation Results}
Table \ref{tab:main} demonstrates the performance of audio beat matching. A time unit is determined as a proper transition location if its corresponding predicted confidence is higher than 0.5. We use this threshold throughout the whole experiment. \emph{hit@0} depicted in Table \ref{tab:main} is the conventional evaluation protocol. As clearly evidenced in experimental results, \emph{BeatX} surpasses the group of \emph{Linear} and \emph{CNN-1d} with a considerable margin, reporting over 7.76\% improvements of the F1-Score on average. This situation suggests the necessity of the self-attention mechanism, \emph{i.e.}, the significance of capturing the adjacent neighbors' character. More specifically, by comparing \emph{BeatX} and \emph{Encoder}, we can further conclude that global attention can further boost the audio beat matching performance since the ABM task requires big-picture thinking, yielding about 7.38\% detection gain, which is also very considerable.

\subsection{Ablations}
\label{ablations}
This section eliminates the following problems: 1). how to determine the best fit of the Gaussian radius $\hat{\sigma}$; 2). verify the effects of label scope mask; 3). investigate the effects of different feature combinations in \emph{BeatX}, \emph{i.e.}, the Mel cepstrum, the short-term energy, and the linear projection; 4). detail the ablations on the hyper-parameters in \emph{BeatX}, for example, the number of self-attention heads, the depth of the self-attention layer, and the intermediate dimension of concatenated features, to name a few.

\vspace{-1em}
\subsubsection{Estimation of the standard deviation $\hat{\sigma}$}
Given the target label $y$ and the corresponding label scope mask $S$, we use the \emph{PN} ratio (positive/negative) to describe whether the positive and negative data are balanced. Formally, we define the \emph{PN} ratio in the following manner:

\begin{equation}
PN=\frac{\sum_{i=1}^{N}S(i, \hat{\sigma})\cdot y_i}{\sum_{i=1}^{N}S(i, \hat{\sigma})\cdot (1-y_i)}
\end{equation}

We aim to select a proper $\hat{\sigma}$ to make the \emph{PN} ratio distribute around 1. To achieve this expectation, we calculate the value of \emph{PN} ratio with $\hat{\sigma}$ starting from 0.6 to 1.3 with a stride of 0.1, and the ablation results are depicted in Table \ref{tab:sigma} in detail. Noting that we calculate the \emph{PN} ratio of each data sample in \emph{AutoMatch} and report their average value.

\setlength{\tabcolsep}{3.5pt}
\vskip -0.05in
\begin{table}[t]
\begin{spacing}{1.4}
\begin{center}
\begin{tabular}{cccc>{\columncolor{gray!25}}ccccc}
\hline
$\hat{\sigma}$ & 0.6 & 0.7 & 0.8 & 0.9 & 1.0 & 1.1 & 1.2 & 1.3 \\ \cline{2-9} 
\emph{PN} ratio & 2.21 & 1.53 & 1.20 & 1.01 & 0.89 & 0.81 & 0.75 & 0.71 \\ \hline
\end{tabular}
\end{center}
\end{spacing}
\vskip -0.08in
\caption{The ablations on the selection of Gaussian radius $\hat{\sigma}$. If not specified, we set 0.9 as the default value of $\hat{\sigma}$.}
\label{tab:sigma}
\vskip -0.2in
\end{table}

\vspace{-0.7em}
\subsubsection{The effects of label scope mask $S$}
\label{ablation_scope}

We compare the recommendation performance with and without using the label scope mask $S$. The detailed results are demonstrated in Table \ref{tab:label_scope}. All methods are evaluated under the setting of \emph{hit@0}, \emph{i.e.}, the conventional evaluation protocol. We highlight the F1-Score with and without the label scope using boldface and underline, respectively. The reported performance gain in the following table is their differences. From Table \ref{tab:label_scope}, we can see that the label scope mask $S$ introduces over 3.06\% improvements at least, evidencing the necessity of data balance in long-tail training data. Each group of the experiment runs twice, and we pick their average performance.

\setlength{\tabcolsep}{5pt}
\begin{table}[t]
\begin{spacing}{1.3}
\begin{center}
\begin{tabular}{l|c|cc>{\columncolor{gray!25}}ccc}
\hline
\emph{Methods} & $S$ & \emph{P} & \emph{R} & \emph{F1} & \emph{ACC} & \emph{F1 Gain}\\\hline
\multirow{2}{*}{\emph{Linear}} & \ding{55} &  64.00 &  37.48 &    \underline{46.01} & 74.15  & / \\
& \ding{51}  & 65.46  &  49.26 &  \textbf{53.82}  & 75.97  & \textbf{\textcolor{red}{$\uparrow$ 7.81}} \\ \hline
\multirow{2}{*}{\emph{CNN-1d}} & \ding{55} & 66.99 &  39.43 &  \underline{47.22}  & 75.04 & /  \\
& \ding{51}  & 67.41  & 46.38  &  \textbf{53.50}  &  75.86 & \textbf{\textcolor{red}{$\uparrow$ 6.28}} \\ \hline
\multirow{2}{*}{\emph{Encoder}} & \ding{55} & 66.90 & 40.45 & \underline{48.97} & 74.55 & / \\
& \ding{51} & 67.10  &  47.29 & \textbf{54.20}  & 77.12 &  \textbf{\textcolor{red}{$\uparrow$ 5.23}}  \\ \hline
\multirow{2}{*}{\emph{BeatX}} & \ding{55} & 65.70 & 57.42 & \underline{58.52} & 74.73 & / \\
 & \ding{51} & 67.21  &  60.36 &  \textbf{61.58}  &  79.95 & \textbf{\textcolor{red}{$\uparrow$ 3.06}} \\ \hline
\end{tabular}
\end{center}
\end{spacing}
\vskip -0.09in
\caption{The ablations on our proposed label scope mask $S$. \ding{55} indicates no label scope is performed while \ding{51} means the method employs label scope mask $S$ for generating adaptive loss weight.}
\label{tab:label_scope}
\end{table}

\vspace{-0.7em}
\subsubsection{Ablation of the different feature combinations}
We further investigate the effects of both manually designed descriptors and the learned adaptive features. In Table \ref{tab:combine}, the \emph{Mel}, \emph{Energy}, and \emph{Raw} indicate the Mel cepstrum, then short-term energy, and the raw audio signal, respectively. All experiments resort to the label scope mask for data balancing. From results depicted in Table \ref{tab:combine}, we notice that solely using the raw signal input can achieve a detection performance (F1-Score) of 55.06\%, which is lower than the groups of Mel cepstrum and short-term energy. The sole energy feature yields 58.12\% F1-Score, surpassing the raw signal by a considerable margin of 3.06\%. Through the feature fusion, the \emph{BeatX} performs better on the validation data, reporting about 61.58\% F1-Score. We argue that this is because the handcraft descriptors can help to prevent the model from shortcut learning.
 
\setlength{\tabcolsep}{5.7pt}
\begin{table}[t]
\begin{spacing}{1.3}
\begin{center}
\begin{tabular}{ccc|cc>{\columncolor{gray!25}}cc}
\hline
\emph{Mel} & \emph{Energy} & \emph{Raw} & \emph{P} & \emph{R} & \emph{F1} & \emph{ACC} \\ \hline
\ding{51}& &  & 69.83 & 53.34 & 57.37 & 77.85 \\
& \ding{51} &     &  66.46 & 58.96  &  \underline{\underline{58.12}}  &   77.92  \\
& & \ding{51} & 69.42 &  47.18 & 55.06   &  76.33   \\\hline
\ding{51} & \ding{51} &  & 68.80 & 53.06 &  59.16 &  78.65  \\
& \ding{51} & \ding{51} & 65.69  &  57.21 & 59.75   &  78.33   \\
\ding{51} & & \ding{51}& 66.67  & 59.61  &  \underline{60.25}  &  78.42   \\ \hline
\ding{51}& \ding{51} & \ding{51} & 67.21  &  60.36 &  \textbf{61.58}  & 79.95  \\ \hline
\end{tabular}
\end{center}
\end{spacing}
\vskip -0.08in
\caption{The ablation on the feature combinations in \emph{BeatX}. We use boldface and underline to highlight the top two models with the best performance. The double underline indicates the most discriminative features among the manual descriptors and deep features. The F1-Score reveals the models' overall performance.}
\label{tab:combine}
\vskip -0.15in
\end{table}

\vspace{-0.7em}
\subsubsection{Ablation on the hyper-parameters of \emph{BeatX}}
The key parameters of \emph{BeatX} are the dimension of projection, the depth of the self-attention module, and the number of attention heads. Concerning the \emph{BeatX} is deployed into practical scenarios, we don't ablate it with very deep attention architecture. The overall results are presented in the following table. The speed is evaluated on a single NVIDIA V100 GPU with Float32 data type. From results depicted in Table \ref{tab:hyper}, we can clearly notice that the inference speed decreases significantly with the increase of attention depth, for example, if we double the depth of attention layers from 2 to 4, the inference speed almost halves from 1873.53 time units per second to 1022.05. By comparison, the inference speed demonstrates less sensitivity with respect to the number of attention heads and the intermediate feature dimension. The quantity of trainable parameters increases dramatically with the expansion of the projection dimension. We use the local attention module with 2 layers and 4 heads by default, \emph{i.e.}, the first group.

\setlength{\tabcolsep}{5.8pt}
\begin{table}[t]
\begin{spacing}{1.3}
\begin{center}
\begin{tabular}{cccc|c>{\columncolor{gray!25}}c}
\hline
\#Proj & \#Head & \#Depth & \#Param & Speed & \emph{F1}\\ \hline
128 & 4 & 2 & 5.10M  & 1873.53 & \underline{61.58} \\
256 & 4 & 2 & 15.15M & 1588.22 & 61.26\\
512 & 4 & 2 & 54.12M & 1434.44& \textbf{62.03} \\\hline
128 & 6 & 2 & 5.10M  & 1726.07 & 61.23\\
128 & 8 & 2 & 5.10M  & 1656.89 & 60.98\\
128 & 12 & 2 & 5.10M & 1635.04 & 61.16\\ \hline
128 & 4 & 4 & 6.49M  & 1022.05 & 60.49\\
128 & 4 & 6 & 7.87M  & 724.40 & 59.87 \\
128 & 4 & 8 & 9.25M  & 582.15 & 58.65 \\ \hline
\end{tabular}
\end{center}
\end{spacing}
\vskip -0.08in
\caption{The ablation on the hyper-parameters of \emph{BeatX}. The speed indicates the number of time units (0.1s) the model processed per second. The selection of the model architecture should consider both the performance and its efficiency.}
\label{tab:hyper}
\vskip -0.15in
\end{table}

\vspace{-0.7em}
\subsubsection{Ablation on the local window size $k$}
We perform ablations on the semi-window size $k$ defined in Section \ref{protocol}. Greater window size yields marginal cost in the generation of local contexts, however, it introduces more computation load within the self-attention module. As shown in Table \ref{tab:window}, using window size as 7 leads to more computation cost than performance improvement, and therefore, we pick $k$ as 5 throughout our experiments. We use the GFLOPs, \emph{i.e.}, the floating-point operations, to measure the computation complexity of the model.

\setlength{\tabcolsep}{7pt}
\vskip -0.04in
\begin{table}[htpb]
\begin{spacing}{1.3}
\begin{center}
\begin{tabular}{l|cc>{\columncolor{gray!25}}ccc}
\hline
$k$ & 1 & 3 & 5 & 7 & 9\\ \hline
\emph{F1} & 58.76 & 60.28 & \underline{\underline{61.58}} & \textbf{62.16} & \underline{61.97}\\
\emph{GFLOPs} & 0.46 & 1.02 & 1.55 & 2.21 & 2.87 \\\hline
\end{tabular}
\end{center}
\end{spacing}
\vskip -0.08in
\caption{The ablation on local window size $k$. Both the F1-Score and the GFLOPs are taken into consideration.}
\label{tab:window}
\vskip -0.1in
\end{table}

\vspace{-0.2em}
\section{Conclusion}
This paper formally defines the audio beat matching (ABM) task and its evaluation protocol, which aims to help neural networks develop human-like editing skills. To empower the following research, we further collect the large-scale \emph{AutoMatch} dataset and propose \emph{BeatX} to serve as the starting point. The presented label scope effectively eliminates the data imbalance issues. We hope this work can introduce more inspiration to computer vision research.

{\small
\bibliographystyle{ieee_fullname}
\bibliography{egbib}

\begin{thebibliography}{10}\itemsep=-1pt

\bibitem{navigation}
Michal Adamkiewicz, Timothy Chen, Adam Caccavale, Rachel Gardner, Preston
  Culbertson, Jeannette Bohg, and Mac Schwager.
\newblock Vision-only robot navigation in a neural radiance world.
\newblock {\em IEEE Robotics and Automation Letters}, 7(2):4606--4613, 2022.

\bibitem{text2live}
Omer Bar-Tal, Dolev Ofri-Amar, Rafail Fridman, Yoni Kasten, and Tali Dekel.
\newblock Text2live: Text-driven layered image and video editing.
\newblock In {\em Computer Vision--ECCV 2022: 17th European Conference, Tel
  Aviv, Israel, October 23--27, 2022, Proceedings, Part XV}, pages 707--723.
  Springer, 2022.

\bibitem{bite_transfer}
Suneel Belkhale, Ethan~K Gordon, Yuxiao Chen, Siddhartha Srinivasa, Tapomayukh
  Bhattacharjee, and Dorsa Sadigh.
\newblock Balancing efficiency and comfort in robot-assisted bite transfer.
\newblock In {\em 2022 International Conference on Robotics and Automation
  (ICRA)}, pages 4757--4763. IEEE, 2022.

\bibitem{lowrank}
Srinadh Bhojanapalli, Chulhee Yun, Ankit~Singh Rawat, Sashank~J. Reddi, and
  Sanjiv Kumar.
\newblock Low-rank bottleneck in multi-head attention models, 2020.

\bibitem{deepLTR}
Fatih Cakir, Kun He, Xide Xia, Brian Kulis, and Stan Sclaroff.
\newblock Deep metric learning to rank.
\newblock In {\em Proceedings of the IEEE/CVF conference on computer vision and
  pattern recognition}, pages 1861--1870, 2019.

\bibitem{pairwise}
Zhe Cao, Tao Qin, Tie-Yan Liu, Ming-Feng Tsai, and Hang Li.
\newblock Learning to rank: from pairwise approach to listwise approach.
\newblock In {\em Proceedings of the 24th international conference on Machine
  learning}, pages 129--136, 2007.

\bibitem{EDCN}
Bo Chen, Yichao Wang, Zhirong Liu, Ruiming Tang, Wei Guo, Hongkun Zheng, Weiwei
  Yao, Muyu Zhang, and Xiuqiang He.
\newblock Enhancing explicit and implicit feature interactions via information
  sharing for parallel deep ctr models.
\newblock In {\em Proceedings of the 30th ACM international conference on
  information \& knowledge management}, pages 3757--3766, 2021.

\bibitem{zero_cross}
Chi-hau Chen.
\newblock {\em Signal processing handbook}, volume~51.
\newblock CRC Press, 1988.

\bibitem{nature_audio}
Haoze Chen and Zhijie Zhang.
\newblock Hybrid neural network based on novel audio feature for vehicle type
  identification.
\newblock {\em Scientific Reports}, 11(1):7648, 2021.

\bibitem{pitch_class}
Ning Chen, J~Stephen Downie, Hai-dong Xiao, and Yu Zhu.
\newblock Cochlear pitch class profile for cover song identification.
\newblock {\em Applied Acoustics}, 99:92--96, 2015.

\bibitem{reachbot}
Tony~G Chen, Becky Miller, Crystal Winston, Stephanie Schneider, Andrew Bylard,
  Marco Pavone, and Mark~R Cutkosky.
\newblock Reachbot: A small robot with exceptional reach for rough terrain.
\newblock In {\em 2022 International Conference on Robotics and Automation
  (ICRA)}, pages 4517--4523. IEEE, 2022.

\bibitem{maskface}
Jiankang Deng, Jia Guo, Xiang An, Zheng Zhu, and Stefanos Zafeiriou.
\newblock Masked face recognition challenge: The insightface track report.
\newblock In {\em Proceedings of the IEEE/CVF International Conference on
  Computer Vision}, pages 1437--1444, 2021.

\bibitem{editing_transfer}
Nathan Frey, Peggy Chi, Weilong Yang, and Irfan Essa.
\newblock Automatic non-linear video editing transfer.
\newblock {\em CoRR}, abs/2105.06988, 2021.

\bibitem{text_talk_editing}
Ohad Fried, Ayush Tewari, Michael Zollh{\"o}fer, Adam Finkelstein, Eli
  Shechtman, Dan~B Goldman, Kyle Genova, Zeyu Jin, Christian Theobalt, and
  Maneesh Agrawala.
\newblock Text-based editing of talking-head video.
\newblock {\em ACM Transactions on Graphics (TOG)}, 38(4):1--14, 2019.

\bibitem{shortcuts}
Robert Geirhos, J{\"o}rn-Henrik Jacobsen, Claudio Michaelis, Richard Zemel,
  Wieland Brendel, Matthias Bethge, and Felix~A Wichmann.
\newblock Shortcut learning in deep neural networks.
\newblock {\em Nature Machine Intelligence}, 2(11):665--673, 2020.

\bibitem{deepfm}
Huifeng Guo, Ruiming Tang, Yunming Ye, Zhenguo Li, and Xiuqiang He.
\newblock Deepfm: a factorization-machine based neural network for ctr
  prediction.
\newblock {\em arXiv preprint arXiv:1703.04247}, 2017.

\bibitem{resnet}
Kaiming He, Xiangyu Zhang, Shaoqing Ren, and Jian Sun.
\newblock Deep residual learning for image recognition.
\newblock In {\em Proceedings of the IEEE conference on computer vision and
  pattern recognition}, pages 770--778, 2016.

\bibitem{gebd_report}
Dexiang Hong, Congcong Li, Longyin Wen, Xinyao Wang, and Libo Zhang.
\newblock Generic event boundary detection challenge at cvpr 2021 technical
  report: Cascaded temporal attention network (castanet).
\newblock {\em arXiv preprint arXiv:2107.00239}, 2021.

\bibitem{sc-transformer++}
Dexiang Hong, Xiaoqi Ma, Xinyao Wang, Congcong Li, Yufei Wang, and Longyin Wen.
\newblock Sc-transformer++: Structured context transformer for generic event
  boundary detection.
\newblock {\em CoRR}, abs/2206.12634, 2022.

\bibitem{gebd_winning}
Hyolim Kang, Jinwoo Kim, Kyungmin Kim, Taehyun Kim, and Seon~Joo Kim.
\newblock Winning the cvpr'2021 kinetics-gebd challenge: Contrastive learning
  approach.
\newblock {\em arXiv preprint arXiv:2106.11549}, 2021.

\bibitem{constrained_control}
Ankur Kapoor, Ming Li, and Russell~H Taylor.
\newblock Constrained control for surgical assistant robots.
\newblock In {\em ICRA}, pages 231--236, 2006.

\bibitem{trans_rnn}
A. Katharopoulos, A. Vyas, N. Pappas, and F. Fleuret.
\newblock Transformers are rnns: Fast autoregressive transformers with linear
  attention.
\newblock In {\em Proceedings of the International Conference on Machine
  Learning (ICML)}, 2020.

\bibitem{adaface}
Minchul Kim, Anil~K Jain, and Xiaoming Liu.
\newblock Adaface: Quality adaptive margin for face recognition.
\newblock In {\em Proceedings of the IEEE/CVF Conference on Computer Vision and
  Pattern Recognition}, pages 18750--18759, 2022.

\bibitem{reformer}
Nikita Kitaev, Lukasz Kaiser, and Anselm Levskaya.
\newblock Reformer: The efficient transformer.
\newblock In {\em International Conference on Learning Representations}, 2020.

\bibitem{human_editing}
Sharath Koorathota, Patrick Adelman, Kelly Cotton, and Paul Sajda.
\newblock Editing like humans: a contextual, multimodal framework for automated
  video editing.
\newblock In {\em Proceedings of the IEEE/CVF Conference on Computer Vision and
  Pattern Recognition}, pages 1701--1709, 2021.

\bibitem{health_care}
Maria Kyrarini, Fotios Lygerakis, Akilesh Rajavenkatanarayanan, Christos
  Sevastopoulos, Harish~Ram Nambiappan, Kodur~Krishna Chaitanya, Ashwin~Ramesh
  Babu, Joanne Mathew, and Fillia Makedon.
\newblock A survey of robots in healthcare.
\newblock {\em Technologies}, 9(1):8, 2021.

\bibitem{sc-transformer}
Congcong Li, Xinyao Wang, Dexiang Hong, Yufei Wang, Libo Zhang, Tiejian Luo,
  and Longyin Wen.
\newblock Structured context transformer for generic event boundary detection.
\newblock {\em CoRR}, abs/2206.02985, 2022.

\bibitem{zcr1}
Tao Li, Mitsunori Ogihara, and Qi Li.
\newblock A comparative study on content-based music genre classification.
\newblock In {\em Proceedings of the 26th annual international ACM SIGIR
  conference on Research and development in informaion retrieval}, pages
  282--289, 2003.

\bibitem{enhancing}
Allen Lin, Ziwei Zhu, Jianling Wang, and James Caverlee.
\newblock Enhancing user personalization in conversational recommenders.
\newblock {\em arXiv preprint arXiv:2302.06656}, 2023.

\bibitem{quantum_dot}
Yang Liu, Fei Han, Fushan Li, Yan Zhao, Maosheng Chen, Zhongwei Xu, Xin Zheng,
  Hailong Hu, Jianmin Yao, Tailiang Guo, et~al.
\newblock Inkjet-printed unclonable quantum dot fluorescent anti-counterfeiting
  labels with artificial intelligence authentication.
\newblock {\em Nature communications}, 10(1):2409, 2019.

\bibitem{DIFM}
Wantong Lu, Yantao Yu, Yongzhe Chang, Zhen Wang, Chenhui Li, and Bo Yuan.
\newblock A dual input-aware factorization machine for ctr prediction.
\newblock In {\em Proceedings of the twenty-ninth international conference on
  international joint conferences on artificial intelligence}, pages
  3139--3145, 2021.

\bibitem{video_authenticity}
Marie-Helen Maras and Alex Alexandrou.
\newblock Determining authenticity of video evidence in the age of artificial
  intelligence and in the wake of deepfake videos.
\newblock {\em The International Journal of Evidence \& Proof}, 23(3):255--262,
  2019.

\bibitem{bigdata}
Vivien Marx.
\newblock The big challenges of big data.
\newblock {\em Nature}, 498(7453):255--260, 2013.

\bibitem{ai_interaction}
Ishaq~Azhar Mohammed.
\newblock The interaction between artificial intelligence and identity and
  access management: an empirical study.
\newblock {\em International Journal of Creative Research Thoughts (IJCRT),
  ISSN}, 2320(2882):668--671, 2021.

\bibitem{solar_energy}
Anthony Ortiz, Dhaval Negandhi, Sagar~R Mysorekar, Shivaprakash~K Nagaraju,
  Joseph Kiesecker, Caleb Robinson, Aditi Khurana, Felipe Oviedo, and Juan
  M.~Lavista Ferres.
\newblock An artificial intelligence dataset for solar energy locations in
  india.
\newblock {\em arXiv preprint arXiv:2202.01340}, 2022.

\bibitem{BankNote}
Felipe Oviedo, Srinivas Vinnakota, Eugene Seleznev, Hemant Malhotra, Saqib
  Shaikh, and Juan~Lavista Ferres.
\newblock Banknote-net: Open dataset for assistive currency recognition.
\newblock {\em https://arxiv.org/pdf/2204.03738.pdf}, 2022.

\bibitem{moviecuts}
Alejandro Pardo, Fabian~Caba Heilbron, Juan~Le{\'o}n Alc{\'a}zar, Ali Thabet,
  and Bernard Ghanem.
\newblock Moviecuts: A new dataset and benchmark for cut type recognition.
\newblock In {\em Computer Vision--ECCV 2022: 17th European Conference, Tel
  Aviv, Israel, October 23--27, 2022, Proceedings, Part VII}, pages 668--685.
  Springer, 2022.

\bibitem{synface}
Haibo Qiu, Baosheng Yu, Dihong Gong, Zhifeng Li, Wei Liu, and Dacheng Tao.
\newblock Synface: Face recognition with synthetic data.
\newblock In {\em Proceedings of the IEEE/CVF International Conference on
  Computer Vision}, pages 10880--10890, 2021.

\bibitem{gebd_discerning}
Ayush~K Rai, Tarun Krishna, Julia Dietlmeier, Kevin McGuinness, Alan~F Smeaton,
  and Noel~E O'Connor.
\newblock Discerning generic event boundaries in long-form wild videos.
\newblock {\em arXiv preprint arXiv:2106.10090}, 2021.

\bibitem{overcome_shortcut}
Piyapat Saranrittichai, Chaithanya~Kumar Mummadi, Claudia Blaiotta, Mauricio
  Munoz, and Volker Fischer.
\newblock Overcoming shortcut learning in a target domain by generalizing basic
  visual factors from a source domain.
\newblock In {\em Computer Vision--ECCV 2022: 17th European Conference, Tel
  Aviv, Israel, October 23--27, 2022, Proceedings, Part XXV}, pages 294--309.
  Springer, 2022.

\bibitem{audio_survey}
Garima Sharma, Kartikeyan Umapathy, and Sridhar Krishnan.
\newblock Trends in audio signal feature extraction methods.
\newblock {\em Applied Acoustics}, 158:107020, 2020.

\bibitem{glu}
Noam Shazeer.
\newblock Glu variants improve transformer, 2020.

\bibitem{deepctrTool}
Weichen Shen.
\newblock Deepctr: Easy-to-use,modular and extendible package of deep-learning
  based ctr models.
\newblock \url{https://github.com/shenweichen/deepctr}, 2017.

\bibitem{auto_transition}
Yaojie Shen, Libo Zhang, Kai Xu, and Xiaojie Jin.
\newblock Autotransition: Learning to recommend video transition effects.
\newblock In Shai Avidan, Gabriel~J. Brostow, Moustapha Ciss{\'{e}},
  Giovanni~Maria Farinella, and Tal Hassner, editors, {\em Computer Vision -
  {ECCV} 2022 - 17th European Conference, Tel Aviv, Israel, October 23-27,
  2022, Proceedings, Part {XXXVIII}}, volume 13698 of {\em Lecture Notes in
  Computer Science}, pages 285--300. Springer, 2022.

\bibitem{efficient_att}
Zhuoran Shen, Mingyuan Zhang, Haiyu Zhao, Shuai Yi, and Hongsheng Li.
\newblock Efficient attention: Attention with linear complexities.
\newblock {\em CoRR}, abs/1812.01243, 2018.

\bibitem{gebd_benchmark}
Mike~Zheng Shou, Stan~Weixian Lei, Weiyao Wang, Deepti Ghadiyaram, and Matt
  Feiszli.
\newblock Generic event boundary detection: A benchmark for event segmentation.
\newblock In {\em Proceedings of the IEEE/CVF International Conference on
  Computer Vision}, pages 8075--8084, 2021.

\bibitem{roformer}
Jianlin Su.
\newblock Roformer: Transformer with rotary position embeddings - zhuiyiai.
\newblock Technical report, 2021.

\bibitem{ai_good}
Mariarosaria Taddeo and Luciano Floridi.
\newblock How ai can be a force for good.
\newblock {\em Science}, 361(6404):751--752, 2018.

\bibitem{PLE}
Hongyan Tang, Junning Liu, Ming Zhao, and Xudong Gong.
\newblock Progressive layered extraction (ple): A novel multi-task learning
  (mtl) model for personalized recommendations.
\newblock In {\em Proceedings of the 14th ACM Conference on Recommender
  Systems}, pages 269--278, 2020.

\bibitem{ai_social}
Nenad Toma{\v{s}}ev, Julien Cornebise, Frank Hutter, Shakir Mohamed, Angela
  Picciariello, Bec Connelly, Danielle~CM Belgrave, Daphne Ezer, Fanny Cachat
  van~der Haert, Frank Mugisha, et~al.
\newblock Ai for social good: unlocking the opportunity for positive impact.
\newblock {\em Nature Communications}, 11(1):2468, 2020.

\bibitem{linformer}
Sinong Wang, Belinda~Z. Li, Madian Khabsa, Han Fang, and Hao Ma.
\newblock Linformer: Self-attention with linear complexity, 2020.

\bibitem{listwise}
Fen Xia, Tie-Yan Liu, Jue Wang, Wensheng Zhang, and Hang Li.
\newblock Listwise approach to learning to rank: theory and algorithm.
\newblock In {\em Proceedings of the 25th international conference on Machine
  learning}, pages 1192--1199, 2008.

\bibitem{mfc}
Min Xu, Ling-Yu Duan, Jianfei Cai, Liang-Tien Chia, Changsheng Xu, and Qi Tian.
\newblock Hmm-based audio keyword generation.
\newblock In {\em Advances in Multimedia Information Processing-PCM 2004: 5th
  Pacific Rim Conference on Multimedia, Tokyo, Japan, November 30-December 3,
  2004. Proceedings, Part III 5}, pages 566--574. Springer, 2005.

\bibitem{home_assistant}
Kimitoshi Yamazaki, Ryohei Ueda, Shunichi Nozawa, Mitsuharu Kojima, Kei Okada,
  Kiyoshi Matsumoto, Masaru Ishikawa, Isao Shimoyama, and Masayuki Inaba.
\newblock Home-assistant robot for an aging society.
\newblock {\em Proceedings of the IEEE}, 100(8):2429--2441, 2012.

\bibitem{zcr2}
Xiaoling Yang, Baohua Tan, Jiehua Ding, Jinye Zhang, and Jiaoli Gong.
\newblock Comparative study on voice activity detection algorithm.
\newblock In {\em 2010 International Conference on Electrical and Control
  Engineering}, pages 599--602, 2010.

\bibitem{riserobots}
Ajmal Zemmar, Andres~M Lozano, and Bradley~J Nelson.
\newblock The rise of robots in surgical environments during covid-19.
\newblock {\em Nature Machine Intelligence}, 2(10):566--572, 2020.

\bibitem{deepface}
Maheen Zulfiqar, Fatima Syed, Muhammad~Jaleed Khan, and Khurram Khurshid.
\newblock Deep face recognition for biometric authentication.
\newblock In {\em 2019 international conference on electrical, communication,
  and computer engineering (ICECCE)}, pages 1--6. IEEE, 2019.

\end{thebibliography}
}

\end{document}